\documentclass[runningheads]{llncs}
\usepackage[T1]{fontenc}
\usepackage{graphicx}
\usepackage{multirow}
\usepackage{booktabs}
\usepackage{xcolor,colortbl}
\definecolor{pshigh}{rgb}{1.0,0.5,0.5}
\definecolor{phigh}{rgb}{1.0,0.6,0.6}
\definecolor{pmiddle}{rgb}{1.0,0.8,0.8}
\definecolor{plow}{rgb}{1.0,0.85,0.85}
\definecolor{pslow}{rgb}{1.0,0.9,0.9}
\definecolor{nshigh}{rgb}{0.7,0.7,1.0}
\definecolor{nhigh}{rgb}{0.75,0.75,1.0}
\definecolor{nmiddle}{rgb}{0.8,0.8,1.0}
\definecolor{nlow}{rgb}{0.85,0.85,1.0}
\definecolor{nslow}{rgb}{0.9,0.9,1.0}

\usepackage{mdframed}

\usepackage{rotating}

\begin{document}
\title{Are Large Language Models Really Effective for Training-Free Cold-Start Recommendation?}
\titlerunning{Are LLMs Really Effective for TFCSR?}
%
\author{Genki Kusano, Kenya Abe, Kunihiro Takeoka}
\authorrunning{Kusano et al.}
%
\institute{NEC Corporation\\
\email{\{g-kusano,abe-kenya,k\_takeoka\}@nec.com}
}
\maketitle              
\begin{abstract}
Recommender systems usually rely on large-scale interaction data to learn from users' past behaviors and make accurate predictions. However, real-world applications often face situations where no training data is available, such as when launching new services or handling entirely new users. In such cases, conventional approaches cannot be applied. This study focuses on training-free recommendation, where no task-specific training is performed, and particularly on \textit{training-free cold-start recommendation} (TFCSR), the more challenging case where the target user has no interactions. Large language models (LLMs) have recently been explored as a promising solution, and numerous studies have been proposed. As the ability of text embedding models (TEMs) increases, they are increasingly recognized as applicable to training-free recommendation, but no prior work has directly compared LLMs and TEMs under identical conditions. We present the first controlled experiments that systematically evaluate these two approaches in the same setting. The results show that TEMs outperform LLM rerankers, and this trend holds not only in cold-start settings but also in warm-start settings with rich interactions. These findings indicate that direct LLM ranking is not the only viable option, contrary to the commonly shared belief, and TEM-based approaches provide a stronger and more scalable basis for training-free recommendation.

\keywords{Cold-start Recommendation \and Training-free Methods \and Large Language Models \and Text Embedding}
\end{abstract}
\section{Introduction}

The goal of recommender systems is to accurately predict which items in a candidate set a target user is likely to prefer. Traditional approaches such as collaborative filtering~\cite{DBLP:conf/sigir/light_gcn_0001DWLZ020,DBLP:conf/www/ncf_HeLZNHC17} assume the availability of large-scale user-item interaction data and achieve strong performance through supervised learning that exploits other users and items. However, in real-world settings, two challenging scenarios frequently arise: the \textit{cold-start} case, where a new user has only a few or no interactions, and the \textit{training-free} case, where no training data is available at all, such as at the launch of a new service. Both have long been recognized as fundamental challenges in recommender systems.

Most prior work has addressed only one of the two main constraints. Some studies focus on cold-start users while leveraging abundant behaviors from other users~\cite{DBLP:conf/cikm/zs_csr_FengPZCL21,DBLP:conf/aaai/zsl_csr_LiJL00H19}, whereas others operate without training data but rely on sufficient behaviors of the target user~\cite{DBLP:conf/ecir/llm_rec_sort_HouZLLXMZ24,DBLP:conf/naacl/WangL24_demo}. As a result, cases with no training data and a few interactions for the target user remain insufficiently studied.

This study focuses on the most restrictive setting, \textit{training-free cold-start recommendation} (TFCSR). We first identify methods that can be applied to TFCSR and then classify them into two approaches. The first uses large language models (LLMs) to produce rankings directly from user and item texts~\cite{DBLP:conf/recsys/llm_rec_listwise_DaiSZYSXS0X23,DBLP:conf/cikm/llm_zsr_HeXJSLFMKM23,DBLP:conf/ecir/llm_rec_sort_HouZLLXMZ24,DBLP:journals/recsys/revisit_prompt_kusano,DBLP:journals/corr/23_is_chatgpt,tois2023_llmrec_Tang2023OneMF,DBLP:conf/naacl/WangL24_demo,DBLP:journals/tkdd/tapping_XuZLWCZW25}. Many studies have suggested that using LLMs directly is considered the only effective option in training-free environments. The second approach uses text embedding models (TEMs), which map user and item texts into a shared vector space where items are ranked by similarity. Compared to classical TEMs such as the BERT series~\cite{DBLP:conf/naacl/bert_DevlinCLT19,DBLP:conf/sigir/colbert_KhattabZ20}, recent TEMs trained with LLM supervision~\cite{DBLP:journals/corr/gte_abs-2308-03281} show strong performance in training-free settings~\cite{DBLP:journals/corr/do_we_need_recsys}.

However, existing studies lack a systematic comparison of LLMs and TEMs under the same training-free conditions, and their effectiveness for cold-start users, especially in the narrow case where no interactions are available, has not been investigated. Several factors explain why this area has been difficult to study. Research on LLM reranking often adopts evaluation designs with candidate set sizes constrained by input length, which differ from those used for TEMs. The high cost and latency of applying LLMs to large candidate sets also limit evaluations at scale. Moreover, strong TEMs trained with LLMs have become widely available only in recent years, meaning that a reliable basis for comparison has been established only very recently. The lack of public datasets with profile information, due to privacy and anonymity issues, and the absence of standard benchmarks have also contributed to this gap. As a result, the training-free and cold-start setting has rarely been the main focus of previous work. To address this, we present the first direct comparison of LLMs and TEMs in TFCSR and provide quantitative evidence of their effectiveness.

The numerical experiments reveal the following results. In most TFCSR cases, {\em TEMs achieve higher accuracy than LLMs used as rerankers}, and a similar tendency is observed in warm-start settings where users have rich interactions, which directly challenges the commonly shared belief that LLMs are always the best choice for training-free environments. Among the TEMs, those trained with LLMs achieve the highest accuracy, suggesting that LLMs are more effective when used to produce embeddings than when applied to direct reranking. These findings provide clear guidance for future research on TFCSR.

\section{Related Work}
\label{sec:related}

\subsubsection{Classification Framework}
\begin{table*}[t]
  \centering
  \caption{Classification of recommendation settings. The taxonomy is defined by whether the model is trained and by the number of interactions of target users $m$.}
  \label{tab:taxonomy}
  \setlength{\tabcolsep}{6pt}
  \begin{tabular}{@{}c|l p{0.65\textwidth} l@{}}
    \toprule
     & \textbf{Size $m$} & \textbf{Description} & \textbf{References} \\
    \midrule
    \multirow[c]{3}{*}{\rotatebox{90}{Need Training}}  
      & Rich & The model is trained with sufficient data from many users, and the target user also has rich interactions. \colorbox{blue!10}{This is a major research focus in recommender systems.} &~\cite{DBLP:conf/sigir/light_gcn_0001DWLZ020,DBLP:conf/www/ncf_HeLZNHC17,DBLP:conf/cikm/bert4rec_SunLWPLOJ19,DBLP:conf/naacl/WangL24_demo} \\ \cmidrule{2-4} 
      & A few & The model is trained, while the target user has only a few interactions (broad cold-start). &~\cite{DBLP:conf/www/tanp_csr_Lin00PCW21,DBLP:conf/www/coldnas_csr_WuWJDDY23} \\ \cmidrule{2-4} 
      & Zero & The model is trained, but the target user has no interactions and only profile information (narrow cold-start). &~\cite{DBLP:conf/cikm/zs_csr_FengPZCL21,DBLP:conf/aaai/zsl_csr_LiJL00H19,DBLP:conf/nips/dropout_VolkovsYP17} \\
    \midrule
     \multirow[c]{3}{*}{\rotatebox{90}{Training-free}} 
      & Rich & No training is performed; recommendations rely only on the target user's own rich interactions. &~\cite{DBLP:conf/recsys/llm_rec_listwise_DaiSZYSXS0X23,DBLP:conf/ecir/llm_rec_sort_HouZLLXMZ24,DBLP:journals/corr/23_is_chatgpt,tois2023_llmrec_Tang2023OneMF} \\ \cmidrule{2-4} 
      & A few & No training is performed; the target user has only a few interactions (broad cold-start). &~\cite{DBLP:conf/cikm/llm_zsr_HeXJSLFMKM23,DBLP:journals/tors/csr_bandit_SilvaSWRP23} \\ \cmidrule{2-4} 
      & Zero & No training is performed; the target user has no interactions and only profile information (narrow cold-start). \colorbox{red!15}{This is the primary focus of our study.} & - \\
    \bottomrule
  \end{tabular}
\end{table*}

To clarify the scope of TFCSR, we propose a classification framework based on two independent axes: whether the model is trained and the number of interactions of the target user (Table~\ref{tab:taxonomy}). For the number of interactions $m$, we define the case with no interactions ($m=0$) as {\em narrow cold-start} (narrow CS), the case with a few interactions (about five or fewer) as {\em broad CS}, and all other cases as the standard warm-start setting. Our analysis focuses on regions where models are training-free and users are in cold-start conditions. In particular, the training-free and narrow CS case has not been directly examined in previous studies. This case represents one of the most challenging yet practically important settings.

\subsubsection{Training-Based Approaches for Cold-Start Recommendation}
Among supervised methods, studies on broad CS have proposed methods that estimate user preference tendencies from a few interactions and adapt globally learned knowledge to individual users~\cite{DBLP:conf/www/tanp_csr_Lin00PCW21,DBLP:conf/www/coldnas_csr_WuWJDDY23}.  In contrast, studies on narrow CS have proposed methods such as generating virtual interactions from user profiles to compensate for the lack of behavioral history~\cite{DBLP:conf/cikm/zs_csr_FengPZCL21}, reconstruct interactions from user attributes using autoencoders~\cite{DBLP:conf/aaai/zsl_csr_LiJL00H19}, and train models with deliberately removed interaction inputs to enable predictions based on side information~\cite{DBLP:conf/nips/dropout_VolkovsYP17}.

\subsubsection{Training-Free Recommendation}
Several studies have explored training-free recommendation methods that construct prompts from a small number of interactions and use LLMs to generate recommendation results~\cite{DBLP:conf/recsys/llm_rec_listwise_DaiSZYSXS0X23,DBLP:conf/cikm/llm_zsr_HeXJSLFMKM23,DBLP:conf/ecir/llm_rec_sort_HouZLLXMZ24,DBLP:journals/recsys/revisit_prompt_kusano,DBLP:journals/corr/23_is_chatgpt,tois2023_llmrec_Tang2023OneMF}.  
In contrast, another line of work~\cite{DBLP:journals/corr/do_we_need_recsys} investigates general text embeddings~\cite{DBLP:journals/corr/gte_abs-2308-03281} as domain-agnostic models and reports that unsupervised TEMs can sometimes outperform domain-specific or fine-tuned models.  

Most of these studies assume nonzero interactions, and their effectiveness in the narrow CS setting has not been systematically examined. To address this gap, our study focuses on TFCSR, which combines the training-free condition with both broad and narrow CS, and provides the first systematic comparison of LLMs and TEMs under the same conditions.

\section{Experiments}

\subsubsection{Datasets}
We use three public datasets: MovieLens-1M~\cite{DBLP:journals/tiis/movielens_HarperK16}, the Job recommendation dataset~\cite{job-recommendation}, and the Amazon Review dataset~\cite{DBLP:journals/corr/amazon_review23}, all of which include natural language metadata such as product names and categories. User profiles are available in MovieLens and the Job dataset. In MovieLens, attributes such as age, gender, and occupation are converted into text. The Job dataset provides more detailed attributes, including university degrees, management experience, and work history. We define two settings: ``Job (exp)'', where the profile includes work experience, and ``Job (no-exp)'', where work history is excluded. Because the Amazon dataset does not contain user profiles for privacy reasons, only the broad CS setting is considered. It is further divided into three categories: Music, Movie, and Book. To ensure comparability, the construction of CS conditions is standardized as described below.

\subsubsection{Problem Formulation}
The recommendation task takes user information $D_u$ as input and ranks a candidate set $I_u=\{t_1,\ldots\}$ according to the target user's preferences. User information can take two forms: (1) profile text $t_u$, such as age or occupation, with no interactions, referred to as the narrow CS case ($D_u=\{t_u\}$); and (2) a small set of $m$ items previously liked by the user, referred to as the broad CS case ($D_u=\{t_{u,1},\dots,t_{u,m}\}$), where each $t_{u,i}$ is the text of item $i$. To illustrate these settings, we provide an example of a user profile and an item (job posting) from the Job dataset\footnote{``Job (no-exp)'' includes only users whose profiles contain no work history.}.

\begin{mdframed}[linewidth=1pt, backgroundcolor=gray!10, roundcorner=5pt]
\noindent\textbf{user profle} ($t_u$): \{'DegreeType': "Bachelor's", 'Major': 'Economics', 'GraduationYear': 2001, 'WorkHistoryCount': 3, 'TotalYearsExperience': 9, 'CurrentlyEmployed': 'Yes', 'ManagedOthers': 'No', 'ManagedHowMany': 0, 'work history': \{'1': 'Customer Service Representative', ...\}\}

\noindent\textbf{item} ($t_i$): \{'title': 'Office Manager/Administrative Assistant', 'Description': '[...] Responsibilities of the Office Manager/Administrative Assistant will include: Scheduling meetings, [...]', 'Requirements': '[...] Candidate must have at least 3 years administrative or office management experience. Excellent proficiency with Word, Excel, and Outlook is required. [...]'\}
\end{mdframed}

To construct the CS settings, we sample 500 users from each dataset who have at least $m+3$ interactions, reserving the most recent three for evaluation. In the broad CS setting ($m>0$), the remaining $m$ items form $D_u$, while in the narrow CS setting ($m=0$) only the profile text is used. The candidate set $I_u$ always contains 50 items: the three reserved positives and 47 randomly sampled unseen items from the same domain but outside the user's history.

\subsection{Methods}

\subsubsection{Sparse Encoder-based Retrieval (BM25)}
As an unsupervised ranking method, we use BM25~\cite{DBLP:journals/ipm/bm25_JonesWR00}. It ranks candidate items based on keyword overlap between the user and item descriptions, where the user text is the profile in the narrow CS setting and the concatenation of interacted items in the broad CS setting.

\subsubsection{Dense Encoder-based Retrieval (TEM)}
We use TEMs to map user text $t_u$ and item text $t_i$ into vectors $\mathbf{v}_u = f(t_u)$ and $\mathbf{v}_i = f(t_i)$. The similarity between two vectors is measured by cosine similarity, $\cos(\mathbf{x}, \mathbf{y}) = \langle \mathbf{x}, \mathbf{y} \rangle / (\|\mathbf{x}\| \|\mathbf{y}\|)$. In the broad CS setting, where the user is represented as $D_u = \{t_{u,1}, \dots, t_{u,m}\}$, the score is computed as the average similarity, $s_{u,i} = m^{-1}\sum_{j=1}^{m} \cos(f(t_{u,j}), \mathbf{v}_i)$.

For the numerical experiments, we use the following TEMs obtained from HuggingFace: \texttt{multilingual-e5-large}~\cite{DBLP:journals/corr/e5_abs-2402-05672}, \texttt{bge-m3}~\cite{DBLP:journals/corr/bge_abs-2402-03216}, \texttt{gte-modernbert-base}, \texttt{gte-Qwen2-1.5B-instruct}, \texttt{gte-Qwen2-7B-instruct}~\cite{DBLP:journals/corr/gte_abs-2308-03281,zhang2024mgte}, \texttt{Qwen3-Embedding-0.6B} and \texttt{Qwen3-Embedding-8B}~\cite{DBLP:journals/corr/qwen3-embedding-abs-2506-05176}\footnote{For \texttt{multilingual-e5-large}, candidate items are embedded as $\mathbf{v}_i = f(\texttt{"passage: "} + t_i)$ and user texts as $\mathbf{v}_u = f(\texttt{"query: "} + t_u)$, following the official instruction. For the Qwen series, candidate items are embedded with the default prefix: \texttt{"Instruct: Given a web search query, retrieve relevant passages that answer the query\textbackslash nQuery: "}.}. The Qwen embedding models were pretrained on synthetic query-document pairs generated by their corresponding LLMs.

\subsubsection{LLM as a Reranker}
For zero-shot inference, we construct a prompt from the user data and candidate items and obtain a ranking as output. The model is instructed to select the top 10 items from 50 candidates and present them in the order most likely preferred by the user. To avoid ordering bias, the presentation order of candidates is randomized each time~\cite{DBLP:conf/ecir/llm_rec_sort_HouZLLXMZ24}. In the experiments, we use \texttt{gpt-4.1-mini} and \texttt{gpt-4.1} from OpenAI, and \texttt{Qwen3-8B}~\cite{DBLP:journals/corr/qwen3-8b_abs-2505-09388} from Alibaba\footnote{The OpenAI models were released on April 14, 2025. \texttt{gpt-4.1-mini} is a cost-efficient model, while \texttt{gpt-4.1} is a stronger but more expensive option. The Qwen embeddings were pretrained with supervision from their corresponding LLMs, thus \texttt{Qwen3-8B} is included for direct comparison with \texttt{Qwen3-Embedding-8B}.}. Survey studies~\cite{DBLP:journals/recsys/revisit_prompt_kusano,DBLP:journals/tkdd/tapping_XuZLWCZW25} report that \texttt{gpt-4.1} is sufficiently strong among current LLMs for recommendation tasks. Based on these studies, we adopted the following prompt template.

\begin{mdframed}[linewidth=1pt, backgroundcolor=gray!10, roundcorner=5pt]
You must solve the following recommendation task. The task is to select 10 items from the candidate set that the user might like and arrange them in order of preference. Output the result as a list consisting of 10 item IDs, like [8, 4, ...]. \{\textbf{user\_info}\}
\texttt{\#} Candidate items: \{1: $t_1$, ... , 50: $t_{50}$\}
\end{mdframed}

Here, \textbf{user\_info} contains the profile in the narrow CS setting and the item history in the broad CS setting.
\begin{mdframed}[linewidth=1pt, backgroundcolor=gray!10, roundcorner=5pt]
\noindent\textbf{user\_info} (narrow CS):
"I am giving you the profile of the target user.
\texttt{\#} User Information: $t_u$"

\noindent\textbf{user\_info} (broad CS):
"I am giving you the items that the target user has interacted with in the past.
\texttt{\#} User Item History (Chronological order, 1 is the oldest):
\{1: $t_{u, 1}$, ..., $m$: $t_{u, m}$\}"
\end{mdframed}

\subsection{Initial Results with LLM and TEM Approaches}
\label{subsec:results}

\begin{table*}[t]
\centering
\caption{Recall@10 and nDCG@10 in the narrow cold-start setting. Statistical significance is evaluated using the one-sided Wilcoxon signed-rank test against \texttt{BM25}; $*$ and $\bigtriangledown$ indicate results significantly higher or lower at $p=10^{-4}$.}
\label{tab:overall_results_profile}
\setlength{\tabcolsep}{4pt}
\scalebox{0.83}{
\begin{tabular}{l|cccccc|cc}
\toprule
 & \multicolumn{2}{c}{ML-1M} & \multicolumn{2}{c}{Job (no exp)} & \multicolumn{2}{c}{Job (exp)} & \multicolumn{2}{|c}{All} \\
 & Recall & nDCG & Recall & nDCG & Recall & nDCG & Recall & nDCG \\
\midrule
\texttt{BM25} & $0.202$ & $0.235$ & $0.321$ & $0.373$ & $0.423$ & $0.476$ & $0.315$ & $0.362$ \\ \midrule
\texttt{gte-modernbert-base} & \cellcolor{pslow}$0.221$ & \cellcolor{pslow}$0.251$ & \cellcolor{nhigh}$0.219^{\bigtriangledown}$ & \cellcolor{nhigh}$0.230^{\bigtriangledown}$ & \cellcolor{nhigh}$0.222^{\bigtriangledown}$ & \cellcolor{nhigh}$0.244^{\bigtriangledown}$ & \cellcolor{nmiddle}$0.221^{\bigtriangledown}$ & \cellcolor{nhigh}$0.242^{\bigtriangledown}$ \\
\texttt{multilingual-e5-large} & \cellcolor{plow}$0.223$ & \cellcolor{plow}$0.260$ & \cellcolor{pslow}$0.343$ & $0.381$ & \cellcolor{nmiddle}$0.335^{\bigtriangledown}$ & \cellcolor{nmiddle}$0.376^{\bigtriangledown}$ & $0.300$ & \cellcolor{nslow}$0.339$ \\
\texttt{bge-m3} & \cellcolor{pslow}$0.220$ & \cellcolor{pslow}$0.249$ & $0.307$ & $0.373$ & \cellcolor{nlow}$0.353^{\bigtriangledown}$ & \cellcolor{nmiddle}$0.371^{\bigtriangledown}$ & \cellcolor{nslow}$0.293$ & \cellcolor{nslow}$0.331$ \\
\texttt{gte-Qwen2-1.5B-instruct} & \cellcolor{plow}$0.233$ & \cellcolor{plow}$0.267$ & \cellcolor{plow}$0.361$ & \cellcolor{pslow}$0.402$ & $0.427$ & $0.462$ & \cellcolor{pslow}$0.340$ & $0.377$ \\
\texttt{gte-Qwen2-7B-instruct} & \cellcolor{pslow}$0.222$ & \cellcolor{pslow}$0.255$ & \cellcolor{plow}$0.382^{*}$ & \cellcolor{plow}$0.440$ & \cellcolor{pmiddle}$0.511^{*}$ & \cellcolor{plow}$0.540^{*}$ & \cellcolor{plow}$0.372^{*}$ & \cellcolor{plow}$0.412^{*}$ \\
\texttt{Qwen3-Embedding-0.6B} & \cellcolor{plow}$0.223$ & \cellcolor{plow}$0.260$ & \cellcolor{pmiddle}$0.397^{*}$ & \cellcolor{plow}$0.439$ & \cellcolor{pslow}$0.465$ & \cellcolor{pslow}$0.503$ & \cellcolor{plow}$0.362^{*}$ & \cellcolor{plow}$0.401$ \\
\texttt{Qwen3-Embedding-8B} & \cellcolor{pmiddle}$0.247$ & \cellcolor{plow}$\textbf{0.273}$ & \cellcolor{pmiddle}$\textbf{0.408}^{*}$ & \cellcolor{pmiddle}$\textbf{0.459}^{*}$ & \cellcolor{pmiddle}$\textbf{0.521}^{*}$ & \cellcolor{plow}$\textbf{0.557}^{*}$ & \cellcolor{pmiddle}$\textbf{0.392}^{*}$ & \cellcolor{plow}$\textbf{0.430}^{*}$ \\ \midrule
\texttt{gpt-4.1-mini} & \cellcolor{pslow}$0.218$ & \cellcolor{nlow}$0.207$ & \cellcolor{nlow}$0.273$ & \cellcolor{nhigh}$0.245^{\bigtriangledown}$ & \cellcolor{nlow}$0.355$ & \cellcolor{nhigh}$0.327^{\bigtriangledown}$ & \cellcolor{nlow}$0.282$ & \cellcolor{nmiddle}$0.260^{\bigtriangledown}$ \\ 
\texttt{gpt-4.1} & \cellcolor{phigh}$\textbf{0.269}^{*}$ & \cellcolor{pslow}$0.247$ & \cellcolor{nlow}$0.283$ & \cellcolor{nhigh}$0.258^{\bigtriangledown}$ & \cellcolor{nslow}$0.381$ & \cellcolor{nmiddle}$0.343^{\bigtriangledown}$ & $0.311$ & \cellcolor{nmiddle}$0.283^{\bigtriangledown}$ \\
\texttt{Qwen/Qwen3-8B} & \cellcolor{nslow}$0.187$ & \cellcolor{nmiddle}$0.165^{\bigtriangledown}$ & \cellcolor{nshigh}$0.061^{\bigtriangledown}$ & \cellcolor{nshigh}$0.057^{\bigtriangledown}$ & \cellcolor{nshigh}$0.092^{\bigtriangledown}$ & \cellcolor{nshigh}$0.080^{\bigtriangledown}$ & \cellcolor{nshigh}$0.114^{\bigtriangledown}$ & \cellcolor{nshigh}$0.101^{\bigtriangledown}$ \\
\bottomrule
\end{tabular}
}
\end{table*}

\begin{table*}[t]
\centering
\caption{Recall@10 and nDCG@10 for the broad cold-start setting with $m=1$.}
\label{tab:overall_results_1-shot}
\setlength{\tabcolsep}{4pt}
\scalebox{0.61}{
\begin{tabular}{l|cccccccccc|cc}
\toprule
 & \multicolumn{2}{c}{ML-1M} & \multicolumn{2}{c}{Job} & \multicolumn{2}{c}{Music} & \multicolumn{2}{c}{Movie} & \multicolumn{2}{c}{Books} & \multicolumn{2}{|c}{All} \\
 & Recall & nDCG & Recall & nDCG & Recall & nDCG & Recall & nDCG & Recall & nDCG & Recall & nDCG  \\
 \midrule
\texttt{BM25} & $0.388$ & $0.423$ & $0.414$ & $0.475$ & $0.345$ & $0.457$ & $0.367$ & $0.442$ & $0.473$ & $0.590$ & $0.397$ & $0.477$ \\ \midrule
\texttt{gte-modernbert-base} & \cellcolor{pslow}$0.426$ & \cellcolor{pslow}$0.446$ & \cellcolor{nlow}$0.371$ & \cellcolor{nslow}$0.441$ & \cellcolor{phigh}$0.498^{*}$ & \cellcolor{pmiddle}$0.561^{*}$ & \cellcolor{pmiddle}$0.456^{*}$ & \cellcolor{plow}$0.519^{*}$ & \cellcolor{plow}$0.557^{*}$ & \cellcolor{pslow}$0.628$ & \cellcolor{plow}$0.462^{*}$ & \cellcolor{pslow}$0.519^{*}$ \\
\texttt{multilingual-e5-large} & \cellcolor{plow}$0.436$ & $0.439$ & \cellcolor{nlow}$0.337^{\bigtriangledown}$ & \cellcolor{nlow}$0.393^{\bigtriangledown}$ & \cellcolor{phigh}$0.464^{*}$ & \cellcolor{plow}$0.543^{*}$ & \cellcolor{plow}$0.439^{*}$ & \cellcolor{plow}$0.488$ & \cellcolor{plow}$0.546^{*}$ & \cellcolor{pslow}$0.631$ & \cellcolor{plow}$0.444^{*}$ & $0.499$ \\
\texttt{bge-m3} & \cellcolor{plow}$0.439$ & $0.442$ & $0.417$ & $0.476$ & \cellcolor{pmiddle}$0.429^{*}$ & \cellcolor{plow}$0.505$ & \cellcolor{plow}$0.423^{*}$ & \cellcolor{plow}$0.491$ & \cellcolor{plow}$0.541^{*}$ & $0.616$ & \cellcolor{plow}$0.449^{*}$ & \cellcolor{pslow}$0.506^{*}$ \\
\texttt{gte-Qwen2-1.5B-instruct} & \cellcolor{plow}$0.459^{*}$ & \cellcolor{plow}$0.473$ & \cellcolor{plow}$0.471^{*}$ & \cellcolor{pslow}$0.517$ & \cellcolor{pshigh}$0.532^{*}$ & \cellcolor{pmiddle}$0.581^{*}$ & \cellcolor{phigh}$0.482^{*}$ & \cellcolor{pmiddle}$0.554^{*}$ & \cellcolor{pmiddle}$0.575^{*}$ & \cellcolor{plow}$0.655^{*}$ & \cellcolor{pmiddle}$0.504^{*}$ & \cellcolor{plow}$0.556^{*}$ \\
\texttt{gte-Qwen2-7B-instruct} & \cellcolor{plow}$0.465^{*}$ & \cellcolor{plow}$0.467$ & \cellcolor{pmiddle}$\textbf{0.511}^{*}$ & \cellcolor{plow}$\textbf{0.554}^{*}$ & \cellcolor{pshigh}$0.535^{*}$ & \cellcolor{phigh}$\textbf{0.604}^{*}$ & \cellcolor{pshigh}$\textbf{0.560}^{*}$ & \cellcolor{phigh}$\textbf{0.598}^{*}$ & \cellcolor{pmiddle}$\textbf{0.609}^{*}$ & \cellcolor{plow}$\textbf{0.683}^{*}$ & \cellcolor{phigh}$\textbf{0.536}^{*}$ & \cellcolor{pmiddle}$\textbf{0.581}^{*}$ \\
\texttt{Qwen3-Embedding-0.6B} & \cellcolor{plow}$\textbf{0.465}^{*}$ & \cellcolor{plow}$0.473$ & \cellcolor{plow}$0.458$ & \cellcolor{pslow}$0.515$ & \cellcolor{pmiddle}$0.445^{*}$ & \cellcolor{plow}$0.523^{*}$ & \cellcolor{pmiddle}$0.447^{*}$ & \cellcolor{plow}$0.515^{*}$ & \cellcolor{plow}$0.553^{*}$ & \cellcolor{pslow}$0.625$ & \cellcolor{plow}$0.474^{*}$ & \cellcolor{plow}$0.530^{*}$ \\
\texttt{Qwen3-Embedding-8B} & \cellcolor{plow}$0.465^{*}$ & \cellcolor{plow}$\textbf{0.477}$ & \cellcolor{plow}$0.475^{*}$ & \cellcolor{plow}$0.533$ & \cellcolor{pshigh}$\textbf{0.535}^{*}$ & \cellcolor{phigh}$0.604^{*}$ & \cellcolor{phigh}$0.509^{*}$ & \cellcolor{phigh}$0.578^{*}$ & \cellcolor{pmiddle}$0.587^{*}$ & \cellcolor{plow}$0.652$ & \cellcolor{pmiddle}$0.514^{*}$ & \cellcolor{plow}$0.569^{*}$ \\ \midrule
\texttt{gpt-4.1-mini} & \cellcolor{nlow}$0.339$ & \cellcolor{nhigh}$0.289^{\bigtriangledown}$ & \cellcolor{nhigh}$0.272^{\bigtriangledown}$ & \cellcolor{nhigh}$0.264^{\bigtriangledown}$ & \cellcolor{plow}$0.405^{*}$ & \cellcolor{nmiddle}$0.365^{\bigtriangledown}$ & \cellcolor{nlow}$0.323$ & \cellcolor{nmiddle}$0.311^{\bigtriangledown}$ & $0.487$ & \cellcolor{nmiddle}$0.427^{\bigtriangledown}$ & \cellcolor{nslow}$0.365$ & \cellcolor{nhigh}$0.331^{\bigtriangledown}$ \\
\texttt{gpt-4.1} & $0.391$ & \cellcolor{nmiddle}$0.336^{\bigtriangledown}$ & \cellcolor{nlow}$0.363$ & \cellcolor{nhigh}$0.328^{\bigtriangledown}$ & \cellcolor{phigh}$0.453^{*}$ & \cellcolor{nlow}$0.396$ & \cellcolor{pslow}$0.399$ & \cellcolor{nlow}$0.357^{\bigtriangledown}$ & \cellcolor{plow}$0.556^{*}$ & \cellcolor{nmiddle}$0.470^{\bigtriangledown}$ & \cellcolor{pslow}$0.433^{*}$ & \cellcolor{nmiddle}$0.377^{\bigtriangledown}$ \\
\texttt{Qwen/Qwen3-8B} & \cellcolor{nmiddle}$0.295^{\bigtriangledown}$ & \cellcolor{nhigh}$0.251^{\bigtriangledown}$ & \cellcolor{nshigh}$0.085^{\bigtriangledown}$ & \cellcolor{nshigh}$0.085^{\bigtriangledown}$ & \cellcolor{nhigh}$0.203^{\bigtriangledown}$ & \cellcolor{nshigh}$0.197^{\bigtriangledown}$ & \cellcolor{nshigh}$0.175^{\bigtriangledown}$ & \cellcolor{nshigh}$0.170^{\bigtriangledown}$ & \cellcolor{nhigh}$0.266^{\bigtriangledown}$ & \cellcolor{nshigh}$0.254^{\bigtriangledown}$ & \cellcolor{nhigh}$0.205^{\bigtriangledown}$ & \cellcolor{nshigh}$0.191^{\bigtriangledown}$ \\
\bottomrule
\end{tabular}
}
\end{table*}

We report the performance of the main methods across all datasets under the narrow CS setting in Table~\ref{tab:overall_results_profile} and under the broad CS setting with $m=1$ in Table~\ref{tab:overall_results_1-shot}. In many cases, LLMs used as rerankers achieved lower accuracy than \texttt{BM25} and were clearly weaker than TEMs such as \texttt{gte-Qwen2-7B-Instruct} and \texttt{Qwen3-Embedding-8B}. For LLMs, \texttt{gpt-4.1} sometimes produced higher Recall than \texttt{BM25}, but the difference in nDCG was not significant. This result aligns with previous studies~\cite{DBLP:conf/emnlp/llm_reranker_Chao0ZL24,DBLP:journals/tkdd/tapping_XuZLWCZW25}, which reported that LLM rerankers can judge item relevance but often fail to generate proper rankings. In contrast, \texttt{Qwen3-Embedding-8B} showed strong performance across datasets, whereas its LLM counterpart \texttt{Qwen3-8B} reached lower accuracy\footnote{In the Job dataset, item descriptions often exceed 500 tokens, while in MovieLens many items contain only about 10 tokens from titles and categories. The difference in input length may have affected LLM performance.}. Among TEMs, newer versions in the Qwen series achieved higher accuracy than earlier ones. These trends were observed in both the narrow CS and the broad CS setting with $m=1$. \texttt{gte-Qwen2-7B-Instruct} and \texttt{Qwen3-Embedding-8B} achieved the highest accuracy, with little difference between them. In contrast, older TEMs without LLM supervision, such as \texttt{gte-modernbert-base}, performed worse than \texttt{BM25} in the Job dataset, showing that not all TEMs are effective. Overall, the results indicate that LLM rerankers offer only limited value in TFCSR, while newer TEMs trained with LLMs provide stronger and more reliable performance.

\subsubsection{Error Analysis}
The previous results raise an important question: why did LLM rerankers perform much worse than TEMs? One hypothesis is that LLMs and TEMs each perform better for different groups of users, and that the proportion of users for whom TEMs were more effective was larger. To investigate the factors behind this performance gap, we conduct an error analysis focusing on user-level score differences between the two models. Specifically, we compute Recall@10 and nDCG@10 for each user with \texttt{gpt-4.1} and \texttt{Qwen3-Embedding-8B}, and record the win/loss relationships of their scores.

\begin{figure}[t]
    \centering
    \includegraphics[width=\linewidth]{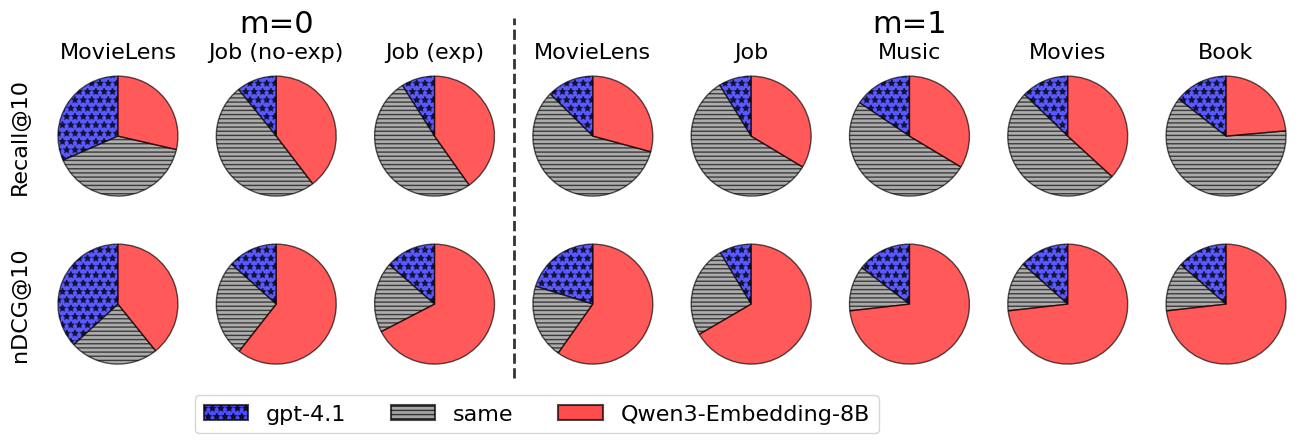}
\caption{Pie charts of user-level win/loss relationships. Each chart shows, for every user, which model obtained the higher score. Users with identical scores are grouped into the ``same'' region.}
    \label{fig:error_analysis}
\end{figure}

Figure~\ref{fig:error_analysis} shows that, except for MovieLens in the narrow CS setting, the proportion of users for whom the LLM obtained higher scores than the TEM was below 25\% in all datasets, indicating that LLMs offered advantages only for a small subset\footnote{For narrow CS in MovieLens, as shown in Table~\ref{tab:overall_results_profile}, the absolute scores were low. Moreover, this setting requires estimating preferences only from age, gender, and occupation, which makes the task especially difficult.}. The TEM outperformed the LLM for a large portion of users in terms of nDCG, while it still maintained an advantage in Recall although the margin was smaller. These results suggest that the weaker performance of LLM rerankers cannot be explained by user-specific characteristics but rather reflects an overall disadvantage compared to TEMs.

\subsection{Additional Analysis}
Based on the above results, this section explores three directions to clarify the limits of LLMs as rerankers and evaluate the potential for improving TEMs.  
(1) Prior studies on LLM rerankers often assume $m\geq5$ interactions and restrict the number of candidate items to about $L=10$ to avoid long prompts. This differs from the TFCSR conditions considered here ($m \leq 1$, $L=50$). We therefore test whether increasing $m$ or reducing $L$ allows LLM rerankers to outperform TEMs.  
(2) TEMs with strong performance were often pretrained using knowledge distilled from LLMs. Another line of research proposes hybrid methods that apply LLMs at inference by transforming raw texts and embedding them. We examine whether such methods can further improve TEM-based approaches.  
(3) As a variant of narrow CS, previous sections considered settings that rely only on user profiles. A related case arises when no interactions are available in the target domain but interactions exist in another domain. We analyze how this cross-domain setting affects the performance of TEMs and LLMs.

\subsubsection{(1) Effect of the Size of User History and Candidate Items}

\begin{figure}[t]
    \centering
    \includegraphics[width=\linewidth]{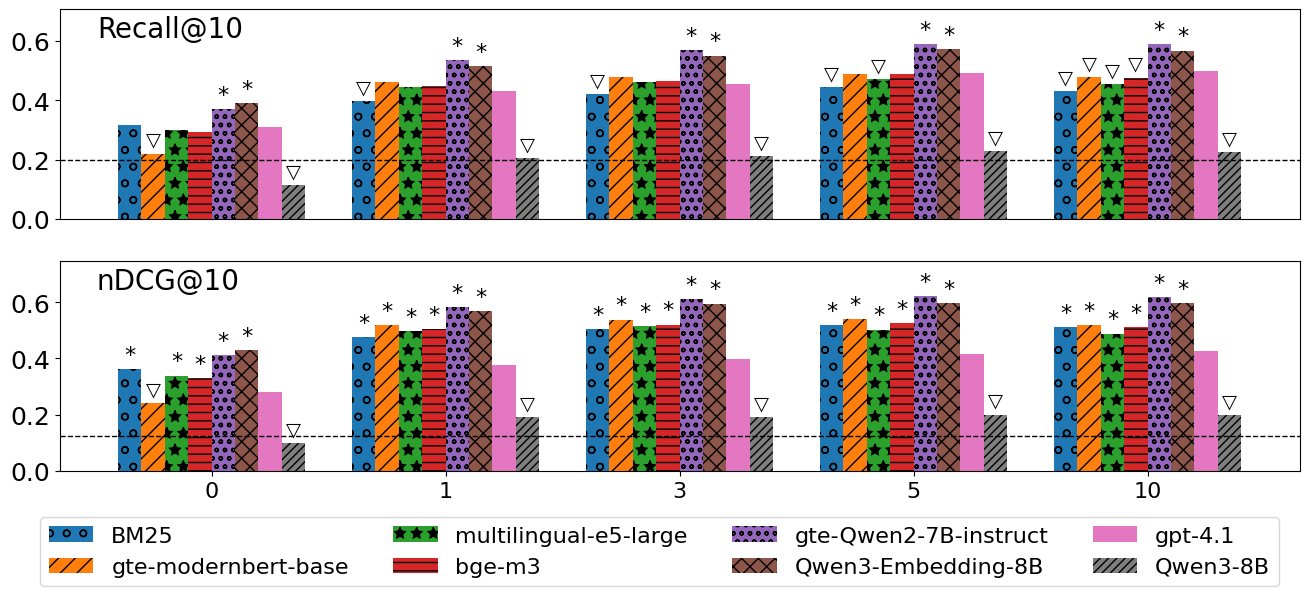}
    \caption{Recall@10 and nDCG@10 for various interaction sizes $m$. Statistical significance is evaluated using the one-sided Wilcoxon signed-rank test against \texttt{gpt-4.1}; $*$ and $\bigtriangledown$ denote results significantly higher or lower at $p=10^{-4}$. The horizontal dotted line represents the score of a random ranking.}
    \label{fig:num_icl}
\end{figure}

We analyze performance trends in the CS setting by varying the number of interactions $m \in \{0,1,3,5,10\}$, where $m=0$ corresponds to the narrow CS. Figure~\ref{fig:num_icl} shows the mean scores across all datasets (corresponding to the ``All'' column in Table~\ref{tab:overall_results_1-shot}). All methods showed improvements as $m$ increased. In Recall, \texttt{gpt-4.1} scored significantly higher than \texttt{BM25} at $m \geq 1$ and \texttt{gte-modernbert-base}, \texttt{multilingual-e5-large}, and \texttt{bge-m3} at $m=10$. However, at $m=10$, it remained significantly weaker than \texttt{gte-Qwen2-7B-Instruct} and \texttt{Qwen3-Embedding-8B}\footnote{While~\cite{DBLP:journals/corr/qwen3-embedding-abs-2506-05176} reported stronger retrieval accuracy for the newer \texttt{Qwen3-Embedding-8B}, in this task \texttt{gte-Qwen2-7B-Instruct} was slightly superior.}. For nDCG, across all $m$, \texttt{gpt-4.1} remained below \texttt{BM25}. These results indicate that adding more interactions improves \texttt{gpt-4.1}, but TEMs also improve, leaving LLMs unable to outperform TEMs.

\begin{figure}[t]
    \centering
    \includegraphics[width=\linewidth]{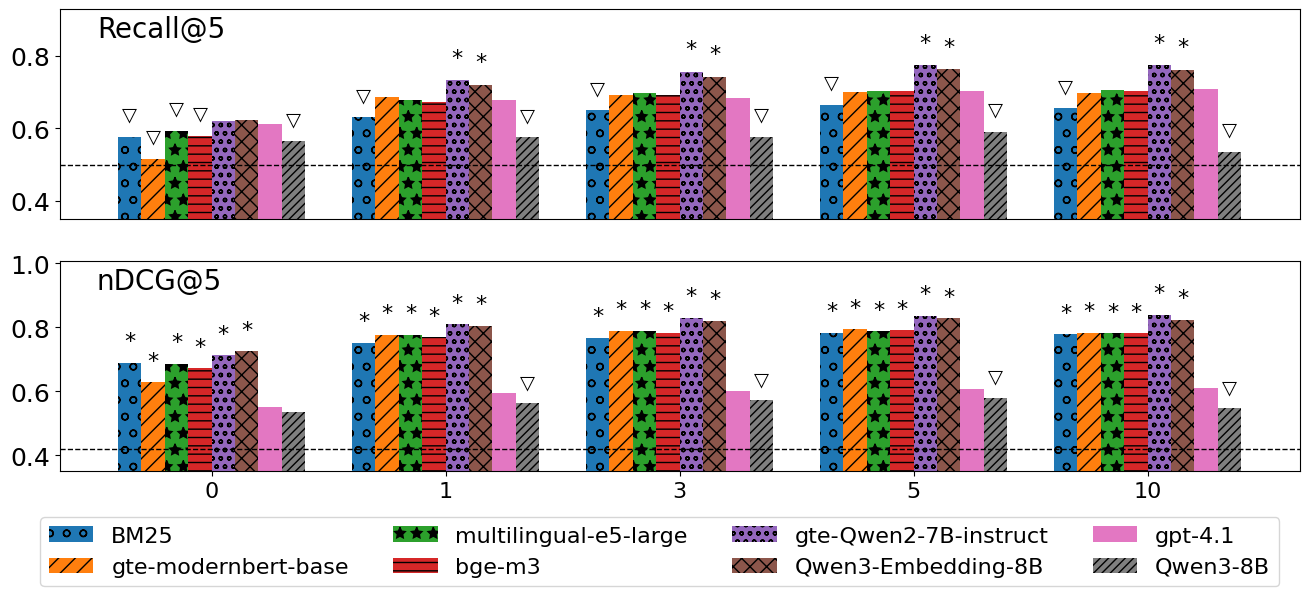}
    \caption{Recall@5 and nDCG@5 when the number of candidate items is 10. The symbols $*$, $\bigtriangledown$, and the horizontal dotted line follow the same definitions as in Figure~\ref{fig:num_icl}.}
    \label{fig:num_icl_small}
\end{figure}

To further examine whether LLMs are affected by long contexts from 50 candidate items, we conducted similar experiments by reducing the candidate set size from 50 to 10. The number of interactions $m$ was varied in the same manner as before. Figure~\ref{fig:num_icl_small} shows that reducing the number of candidate items did not change the overall trend: LLMs still failed to outperform TEMs. When using \texttt{gpt-4.1} as the reference in a Wilcoxon test, the difference in Recall at $m=0$ (narrow CS) was no longer statistically significant, and the performance was comparable to that of the Qwen embedding models. However, for $m \geq 1$, the TEMs achieved higher scores than the LLMs. These results indicate that TEMs remain strong in training-free environments, regardless of changes in the number of interactions and candidate items.

\subsubsection{(2) Effect of Hybrid Methods}
Several methods~\cite{DBLP:journals/corr/querec,DBLP:conf/naacl/llmrec_qeury_expansion_LyuJZXWZCLTL24,DBLP:conf/recsys/kar_XiLLCZZCT0024,DBLP:conf/coling/ur4rec_ZhangZD25} use an LLM to generate query representations, which are then embedded, instead of embedding the raw text directly. Following this approach, we design a query expansion prompt on the user side, based on~\cite{DBLP:journals/corr/querec}, as shown below.

\begin{mdframed}[linewidth=1pt, backgroundcolor=gray!10, roundcorner=5pt]
\noindent\textbf{prompt}: "I am planning to make a recommender system, so please enrich the following user's information.
\texttt{\#} User profile
$t_u$
\texttt{\#} Task
The items to recommend are in the job domain. 
Please generate 10 distinct and comprehensive search queries.

\noindent\textbf{output}: "entry level economics jobs", "customer service jobs", ...
\end{mdframed}

Each user is associated with $K_u=10$ queries. A similar procedure can be applied to items, producing $K_i$ queries for each item. This gives a set of user vectors $\{\mathbf{v}_{u,k}\}_{k=1}^{K_u}$ and item vectors $\{\mathbf{v}_{i,\ell}\}_{\ell=1}^{K_i}$. We then define two types of similarity between user $u$ and candidate item $i$. (\textbf{Max-Sum}) $s_{u,i} = \sum_{k=1}^{K_u} \max_{\ell} \cos(\mathbf{v}_{u,k}, \mathbf{v}_{i,\ell})$, following ColBERT~\cite{DBLP:conf/sigir/colbert_KhattabZ20}. This approach emphasizes the best-matching pairs and ignores weaker alignments. (\textbf{Earth Mover's Distance (EMD)}) This measures the minimum cost of moving one set of vectors into another by treating them as weighted distributions. The cost reflects how efficiently user and item queries can be aligned. The final score is defined as $s_{u,i}=1-w$, where $w$ is the optimal transport cost\footnote{For details of EMD, please refer to \cite{DBLP:conf/icml/wmd_KusnerSKW15}. In experiments, we used the POT library~\cite{DBLP:journals/jmlr/pot_FlamaryCGABCCCF21} with the Sinkhorn algorithm and a regularization parameter $\lambda=10^{-4}$.}. In short, Max-Sum focuses on the strongest local matches, while EMD evaluates the global alignment.

We compare three vector representations. \textbf{Raw} embeds the original text $t$ as a single vector $\mathbf{v}=f(t)$. \textbf{CQ} concatenates multiple queries into one text $t'$ before embedding. \textbf{MQ} embeds each query separately, producing a set $\{\mathbf{v}_k\}$. These forms are combined on the user and item sides. \textbf{Raw-Raw} uses original embeddings. \textbf{CQ-CQ} applies concatenation on both sides, which is equivalent to~\cite{DBLP:journals/corr/querec}. \textbf{CQ-Raw} expands only the user side, while \textbf{MQ-Raw} represents the user with multiple queries and leaves items unchanged. \textbf{MQ-MQ} applies MQ to both and compares the sets using a set-set similarity.

\begin{figure}[t]
    \centering
    \includegraphics[width=\linewidth]{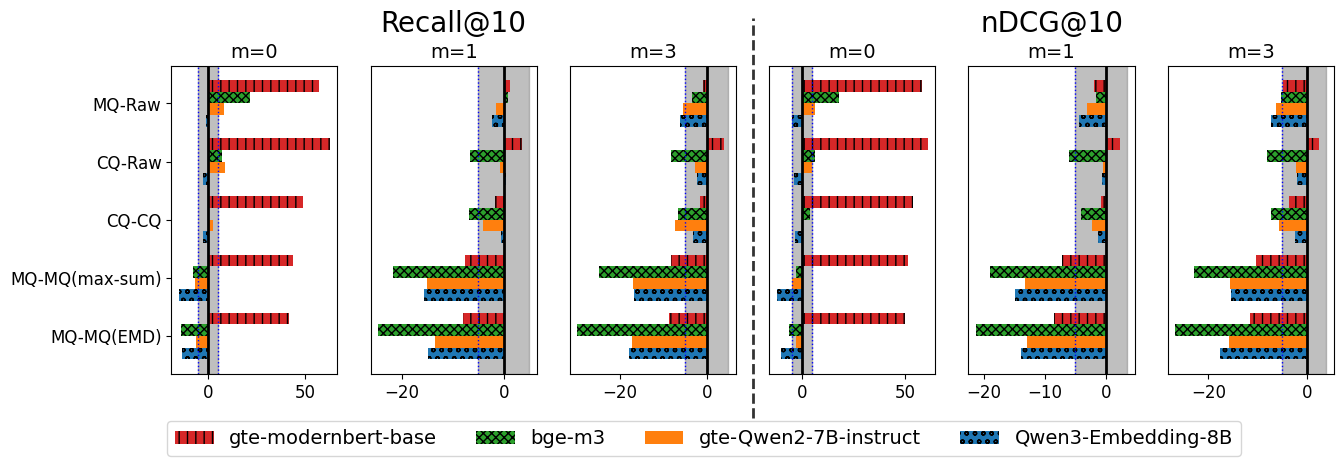}
    \caption{Relative improvement (\%) over \textbf{Raw-Raw} for different TEMs. The light gray area marks the range from $-5\%$ to $5\%$.}
    \label{fig:rq_embedding_raw-raw}
\end{figure}

Figure~\ref{fig:rq_embedding_raw-raw} shows the relative improvement over \textbf{Raw-Raw} when \texttt{gpt-4.1} is used for query expansion. At $m=0$, \texttt{gte-modernbert-base} showed the largest increase, reaching nearly 50\%. Most other TEMs also achieved positive improvements with \textbf{MQ-Raw} and \textbf{CQ-Raw}, except for \texttt{Qwen3-Embedding-8B}. However, the baseline score of \texttt{gte-modernbert-base} with \textbf{Raw-Raw} was originally low, and the improvements of many models with \textbf{MQ-Raw} and \textbf{CQ-Raw} were below 5\% and remained lower than the \textbf{Raw-Raw} of \texttt{Qwen3-Embedding-8B} (see also Table~\ref{tab:overall_results_profile}). When candidate items were also expanded as in \textbf{CQ-CQ} and \textbf{MQ-MQ}, no clear improvement was observed at $m=0$ except for \texttt{gte-modernbert-base}. Moreover, for $m=1$ and $m=3$, almost all expansion methods produced lower scores, suggesting that query expansion was not effective under these conditions. This trend is consistent with~\cite{DBLP:conf/eacl/query_expansion_fail_WellerLWLDCS24}, which found that models with high retrieval accuracy often show lower scores after query expansion, whereas weaker models sometimes improve. In summary, the results indicate that TEMs achieve higher accuracy when user and item texts are kept in their original form without expansion.

\subsubsection{(3) Effect of Cross-Domain Profile} 
This section analyzes a cross-domain scenario. We select two Amazon domains and extract users who appear in both. For each user, three items are set aside from one domain for evaluation, while $m' \in \{1,3,5,10\}$ items from the other provide the inputs for inference. We focus on the Music-Movie and Movie-Book pairs, which are often examined in prior studies on cross-domain recommendation~\cite{DBLP:journals/csur/cdr_KhanIG17,DBLP:conf/wsdm/cdr_ZhuTLZXZLH22}.

\begin{table*}[t]
\centering
\caption{Recall@10 and nDCG@10 in the cross-domain setting with $m'=1$. ``Music $\rightarrow$ Movie'' indicates that the source domain (user behaviors) is Music and the target domain (candidate items) is Movie.}
\label{tab:cross-domain-1}
\setlength{\tabcolsep}{4pt}
\scalebox{0.7}{
\begin{tabular}{l|cccccccc|cc}
\toprule
 & \multicolumn{2}{c}{Music $\rightarrow$ Movie} & \multicolumn{2}{c}{Movie $\rightarrow$ Music} & \multicolumn{2}{c}{Books $\rightarrow$ Movie} & \multicolumn{2}{c}{Movie $\rightarrow$ Books} & \multicolumn{2}{|c}{All}  \\
 & Recall & nDCG & Recall & nDCG & Recall & nDCG & Recall & nDCG & Recall & nDCG \\
\midrule
\texttt{BM25} & $0.249$ & $0.308$ & $0.175$ & $0.214$ & $0.238$ & $0.282$ & $0.197$ & $0.232$ & $0.215$ & $0.259$ \\ \midrule
\texttt{gte-modernbert-base} & \cellcolor{phigh}$0.327^{*}$ & \cellcolor{pmiddle}$0.380$ & \cellcolor{pshigh}$0.307^{*}$ & \cellcolor{pshigh}$0.349^{*}$ & \cellcolor{plow}$0.277$ & \cellcolor{plow}$0.313$ & \cellcolor{phigh}$0.293^{*}$ & \cellcolor{phigh}$0.323^{*}$ & \cellcolor{phigh}$0.301^{*}$ & \cellcolor{phigh}$0.341^{*}$ \\
\texttt{multilingual-e5-large} & \cellcolor{phigh}$0.325^{*}$ & \cellcolor{plow}$0.358$ & \cellcolor{pshigh}$0.267^{*}$ & \cellcolor{phigh}$0.311^{*}$ & \cellcolor{pmiddle}$0.305^{*}$ & \cellcolor{plow}$0.331$ & \cellcolor{phigh}$0.258^{*}$ & \cellcolor{pmiddle}$0.287$ & \cellcolor{phigh}$0.289^{*}$ & \cellcolor{pmiddle}$0.322^{*}$ \\
\texttt{bge-m3} & \cellcolor{plow}$0.293$ & \cellcolor{plow}$0.342$ & \cellcolor{phigh}$0.241^{*}$ & \cellcolor{phigh}$0.283^{*}$ & \cellcolor{plow}$0.266$ & \cellcolor{plow}$0.314$ & \cellcolor{phigh}$0.261^{*}$ & \cellcolor{pmiddle}$0.285$ & \cellcolor{pmiddle}$0.265^{*}$ & \cellcolor{plow}$0.306^{*}$ \\
\texttt{gte-Qwen2-1.5B-instruct} & \cellcolor{phigh}$0.345^{*}$ & \cellcolor{pmiddle}$0.388^{*}$ & \cellcolor{pshigh}$0.331^{*}$ & \cellcolor{pshigh}$\textbf{0.369}^{*}$ & \cellcolor{phigh}$0.351^{*}$ & \cellcolor{phigh}$0.393^{*}$ & \cellcolor{phigh}$0.291^{*}$ & \cellcolor{phigh}$0.310^{*}$ & \cellcolor{pshigh}$0.330^{*}$ & \cellcolor{phigh}$0.365^{*}$ \\
\texttt{gte-Qwen2-7B-instruct} & \cellcolor{pshigh}$\textbf{0.416}^{*}$ & \cellcolor{phigh}$\textbf{0.456}^{*}$ & \cellcolor{pshigh}$\textbf{0.332}^{*}$ & \cellcolor{pshigh}$0.367^{*}$ & \cellcolor{pshigh}$\textbf{0.409}^{*}$ & \cellcolor{pshigh}$\textbf{0.435}^{*}$ & \cellcolor{pshigh}$\textbf{0.321}^{*}$ & \cellcolor{pshigh}$\textbf{0.348}^{*}$ & \cellcolor{pshigh}$\textbf{0.369}^{*}$ & \cellcolor{pshigh}$\textbf{0.402}^{*}$ \\
\texttt{Qwen3-Embedding-0.6B} & \cellcolor{phigh}$0.331^{*}$ & \cellcolor{pmiddle}$0.378$ & \cellcolor{phigh}$0.258^{*}$ & \cellcolor{phigh}$0.293^{*}$ & \cellcolor{phigh}$0.316^{*}$ & \cellcolor{pmiddle}$0.365^{*}$ & \cellcolor{pmiddle}$0.240$ & \cellcolor{pslow}$0.252$ & \cellcolor{phigh}$0.286^{*}$ & \cellcolor{pmiddle}$0.322^{*}$ \\
\texttt{Qwen3-Embedding-8B} & \cellcolor{pshigh}$0.403^{*}$ & \cellcolor{phigh}$0.445^{*}$ & \cellcolor{pshigh}$0.317^{*}$ & \cellcolor{pshigh}$0.349^{*}$ & \cellcolor{pshigh}$0.381^{*}$ & \cellcolor{pshigh}$0.426^{*}$ & \cellcolor{phigh}$0.285^{*}$ & \cellcolor{phigh}$0.311^{*}$ & \cellcolor{pshigh}$0.347^{*}$ & \cellcolor{phigh}$0.383^{*}$ \\ \midrule
\texttt{gpt-4.1-mini} & \cellcolor{nlow}$0.218$ & \cellcolor{nmiddle}$0.230^{\bigtriangledown}$ & \cellcolor{nslow}$0.165$ & \cellcolor{nhigh}$0.147^{\bigtriangledown}$ & \cellcolor{nslow}$0.217$ & \cellcolor{nmiddle}$0.210^{\bigtriangledown}$ & \cellcolor{nmiddle}$0.143$ & \cellcolor{nhigh}$0.122^{\bigtriangledown}$ & \cellcolor{nlow}$0.186$ & \cellcolor{nhigh}$0.177^{\bigtriangledown}$ \\
\texttt{gpt-4.1} & \cellcolor{pslow}$0.265$ & \cellcolor{nlow}$0.262$ & \cellcolor{plow}$0.209$ & \cellcolor{nlow}$0.192$ & \cellcolor{pmiddle}$0.302^{*}$ & $0.271$ & $0.199$ & \cellcolor{nmiddle}$0.171^{\bigtriangledown}$ & \cellcolor{plow}$0.244^{*}$ & \cellcolor{nlow}$0.224^{\bigtriangledown}$ \\
\texttt{Qwen3-8B} & \cellcolor{nshigh}$0.111^{\bigtriangledown}$ & \cellcolor{nshigh}$0.104^{\bigtriangledown}$ & \cellcolor{nshigh}$0.086^{\bigtriangledown}$ & \cellcolor{nshigh}$0.078^{\bigtriangledown}$ & \cellcolor{nhigh}$0.121^{\bigtriangledown}$ & \cellcolor{nshigh}$0.111^{\bigtriangledown}$ & \cellcolor{nshigh}$0.082^{\bigtriangledown}$ & \cellcolor{nshigh}$0.070^{\bigtriangledown}$ & \cellcolor{nshigh}$0.100^{\bigtriangledown}$ & \cellcolor{nshigh}$0.091^{\bigtriangledown}$ \\
\bottomrule
\end{tabular}
}
\end{table*}

\begin{figure}[t]
    \centering
    \includegraphics[width=\linewidth]{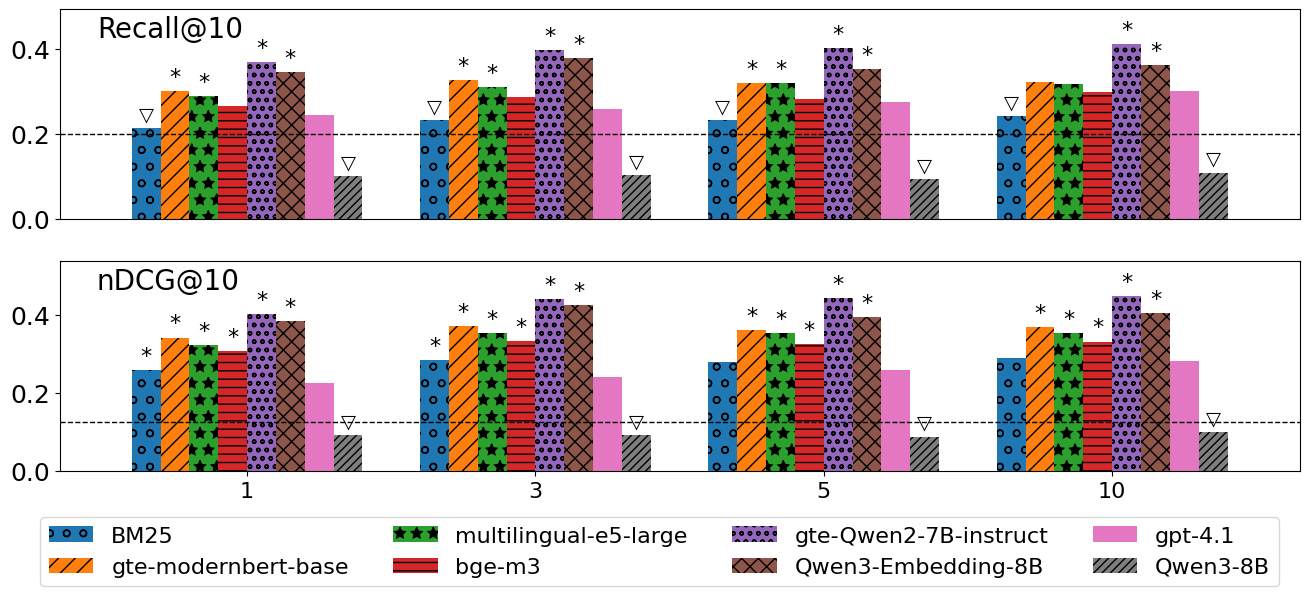}
    \caption{Recall@10 and nDCG@10 for the cross-domain setting with various $m'$.}
    \label{fig:num_icl_cross-domain}
\end{figure}

Table~\ref{tab:cross-domain-1} shows the scores when $m'=1$ is used for inference, as in Table~\ref{tab:overall_results_1-shot}. Figure~\ref{fig:num_icl_cross-domain} presents the average scores across all datasets, following the same procedure as in Figure~\ref{fig:num_icl}. The results show that the Qwen embedding models achieved higher scores than \texttt{gpt-4.1} in all settings. This suggests that TEMs remain the most effective approach even for training-free cross-domain recommendation.

\section{Discussions and Conclusion}
The experiments showed that Qwen embedding models, pretrained on synthetic data generated by their corresponding LLMs, achieved higher accuracy than other TEMs and LLM rerankers. This suggests that LLM supervision is a promising direction for TFCSR. It enables the construction of large pseudo-datasets without real user data and allows TEMs to learn diverse semantic relations even under cold-start conditions.

Future work should address several challenges. Synthetic data to train TEMs may contain domain biases or hallucinated content that limit generalization. Retrieving actual data from the web could be another option, although website terms of service often restrict such use. Moreover, recommendation tasks often involve structured inputs such as profiles or item attributes, while current LLM supervision focuses on natural language. Training methods that integrate structured features and systematic studies on the scale and quality of synthetic data will be necessary to advance this approach.

Although this study focused on the training-free setting, there remains room for further investigation in supervised scenarios. In particular, for cold-start recommendation where training data are available, several approaches have been proposed in which semantic embeddings are used for cold-start items at the beginning, and collaborative information is gradually integrated as training progresses~\cite{DBLP:journals/corr/let_it_go,DBLP:conf/mm/clcrec_WeiWLNLLC21,DBLP:conf/sigir/heater_ZhuSSC20}. These studies rely on relatively old embeddings such as TFIDF, Sentence2Vec~\cite{DBLP:conf/iclr/sentence2vec_AroraLM17}, and \texttt{multilingual-e5}. Replacing them with modern TEMs trained under LLM supervision is expected to provide further improvements.

In conclusion, the results indicate that TEMs trained with LLMs can be more effective than direct reranking with LLMs for TFCSR.

\clearpage

\bibliographystyle{splncs04}
\bibliography{references}

\end{document}